\documentclass[twocolumn,secnumarabic, nobibnotes, aps, prd]{revtex4}
\usepackage{amssymb}

\input{tcilatex}

\begin{document}

\title{Respective influences of pair breaking and phase fluctuations in
disordered high $T_{c}$ superconductors}
\author{F.\ Rullier-Albenque$^{1}$, H.\ Alloul$^{2}$, R.\ Tourbot$^{1}$}
\affiliation{$^{1\text{ }}$SPEC, Orme des Merisiers, CEA, 91191 Gif sur Yvette cedex,
France. \\
$^{2}$ Laboratoire de Physique des Solides, UMR 8502, Universit\'{e}
Paris-Sud, 91405 Orsay, France. }
\date{\today}

\begin{abstract}
Electron irradiation has been used to introduce point defects in a
controlled way in the CuO$_{2}$ planes of underdoped and optimally doped
YBCO.\ This technique allows us to perform very accurate measurements of $%
T_{c}$ and of the residual resistivity in a wide range of defect contents $%
x_{d}$ down to $T_{c}=0$.\ The $T_{c}$ decrease does not follow the
variation expected from pair breaking theories.\ The evolutions of $T_{c}$
and of the transition width with $x_{d}$ emphasize the importance of phase
fluctuations, at least for the highly damaged regime.\ These results open
new questions about the evolution of the defect induced $T_{c}$ depression
over the phase diagram of the cuprates.
\end{abstract}

\maketitle

The establishment of the condensed state in high-$T_{c}$ superconductors
(HTSC) is not fully understood. The occurence of the pseudogap in the
underdoped part of the phase diagram \cite{Timusk} has raised the question
of the coexistence of competing order parameters, or of the occurence of
preformed pairs, with an eventual condensation of the pairs in a coherent
superconducting state at $T_{c}$.\ It has been proposed that, in these low
carrier concentration systems, $T_{c}$ could be determined by the phase
stiffness of the order parameter \cite{Emery1}. The phase fluctuations could
explain the occurence of a direct transition between superconducting and
insulating states.\ An experimental approach which has been used extensively
has been to study the influence of the disorder on both the normal and
superconducting properties.\ In plane impurity substitutions induce T$_{c}$
depression \cite{Chien,Fuku}, modification of the superfluid density $n_{s}$ %
\cite{Nachumi,Bernhard} and local depression of the order parameter \cite%
{STM-Zn}.\ Whether the T$_{c}$ decrease results from pair breaking effects
of the d-wave order parameter \cite{Radtke,Borkowski}, from the
inhomogeneity of the order parameter (so called ''swiss cheese model'' \cite%
{Nachumi}) or from a reduction of the phase stiffness \cite{Emery2} is still
highly debated.\emph{\ }

In order to bring more accurate information on these issues we have
undertaken careful studies of the influence of controlled disorder in single
crystals of HTSC\ cuprates.\ This can be achieved by high energy electron
irradiations performed at low temperatures which introduce only point
defects such as Cu and O vacancies in the CuO$_{2}$\ planes \cite{Legris,
Giapin, Tolpygo}.\ The fact that a unique single crystal is progressively
damaged allows us to study the influence of disorder with an accuracy
impossible to attain with chemical substitutions.\ This gives us the
opportunity to study precisely the properties of samples with highly reduced 
$T_{c}$, as we can indeed control the irradiation fluence to reach $T_{c}=0$%
.\ The results reported in this paper display remarquably simple
phenomenological variations.\ We observe that $T_{c}\;$quite unexpectedly
decreases linearly with defect content down to $T_{c}$ $=0$, in both
underdoped and optimally doped YBCO, at variance with Abrikosov Gorkov based
theories.\ The analysis of the variation of $T_{c}$ and of the transition
width $\delta T_{c}$ reveals that the resistivity at $T_{c}$ is the relevant
parameter which determines the$\;T_{c}$ depression when $T_{c}$ approaches
zero. This agrees with the expected influence of quantum phase
fluctuations.\ We shall discuss whether this approach can describe as well
the variation with hole doping of the initial decrease of $T_{c}$ with
increasing disorder.\ 

The single crystals of YBCO are similar to those studied in \cite{RA}.
In-plane resistivities were measured by the Van der Pauw method.\ Special
care has been taken to put the electrical contacts on the edges of the
samples to insure an homogeneous flow of the electrical current through the
samples. In the case of YBCO$_{7}$, two different crystals of the same batch
( \#1 and\#2) have been studied \cite{footnote1}. The irradiations were
performed with 2.5MeV electrons in the low temperature facility of the Van
der Graaff accelerator at the LSI (Ecole Polytechnique, Palaiseau).\ The
samples were immersed in liquid H$_{2}$ and the electron flux was limited to
10$^{14}e/cm^{2}/s$ \ to avoid heating of the samples during irradiation.
The sample thicknesses ($\simeq 20\mu m$) were much smaller than the
penetration depth of the electrons, which ensured a homogeneous damage
throughout the samples \cite{footnote2}.

$\;$For all samples we have found that Matthiessen's rule is well verified
at high temperature, as the high $T$ parts of the $\rho (T)$ curves shift
parallel to each other.\ This is exemplified in fig.1, which displays the $T$
dependences of the in-plane resistivity $\rho _{ab}$ \ for YBCO$_{7}$ \#2$.$
This confirms that even for very high defect content ($x_{d\text{ }}$\symbol{%
126}9\% in the planes \cite{footnote3}) the hole doping is not significantly
modified as was already shown before for low $x_{d\text{ }}$ \cite{Legris,RA}%
\textit{.}

\FRAME{ftbphFU}{9.0325cm}{6.524cm}{0pt}{\Qcb{The resistivity is plotted
versus temperature for the single crystal YBCO$_{7}$ \#2 irradiated at low
temperature by 2.5 MeV electrons. Data taken either after annealing at 150K
(curve 6) or 300K\ (curve 7) are shown to coincide. }}{\Qlb{Fig.1}}{Figure}{%
\special{language "Scientific Word";type "GRAPHIC";maintain-aspect-ratio
TRUE;display "USEDEF";valid_file "T";width 9.0325cm;height 6.524cm;depth
0pt;original-width 536pt;original-height 426.625pt;cropleft "0";croptop
"1";cropright "1.1052";cropbottom "0";tempfilename
'H8569Y04.wmf';tempfile-properties "XPR";}}

For low $T_{c}$ samples upturns of the resistivity are disclosed\ at low $T$
and increase with increasing $x_{d\text{ }}$. These $"\ln T"$ contributions
have been analysed in the case of YBCO$_{6.6\text{ }}$\cite{RA2} as being
associated with a combination of single impurity scattering and\
localization effects. The latter become dominant for defect contents for
which superconductivity is fully suppressed.\ 

To compare our results with pair breaking theories, it is useful to study
the variation of $T_{c}$ with defect content $x_{d}$.\ The $T_{c}$ values
reported hereafter were measured at the middle point of the resistive
superconducting transition and the error bars were determined from the
10\%-90\% values of the extrapolated normal state resistivity. As we shall
see later on, the sharpness of the resistive transition is a good indication
that the homogeneity of the damage in the sample is ensured.\ 

The best estimate for $x_{d}$ is a priori the irradiation fluence \cite%
{footnote4}.\ For low $x_{d}$, data were taken in the irradiation set-up
without annealing the samples above 150K. We found that $T_{c}$ and the
normal state resistivity $\rho _{ab}$ (measured at 150K) vary linearly with
the irradiation fluence, so that $\Delta \rho _{ab}=\rho _{ab}(150K)-\rho
_{ab}^{pure}(150K)\;$represents $x_{d}$. \ In many cases the samples had to
be taken to room $T$ between irradiation runs.\ Although part of the defects
were annealed in such processes, the $\rho \left( T\right) $ curves were
found to superimpose to those obtained before annealing (\textit{e.g.}
curves 6\ and 7 in fig.1). Therefore $\Delta \rho _{ab\;}$remains a good
estimate for $x_{d}$.\ However, in YBCO$_{6.6}$, \ the variation with
irradiation fluence of $\Delta \rho _{ab},$ measured without annealing above
150K, departs from linearity for $\Delta \rho _{ab}>100\mu \Omega .cm$,
although $T_{c}$ still decreases linearly with the electron fluence. In this
case $\Delta \rho _{ab}$\ has then been replaced by $\Delta \rho _{ab}^{\ast
},$ its linear extrapolation with fluence. We have therefore reported in
figure 2 the data for $T_{c}$ versus both $\Delta \rho _{ab}$ and $\Delta
\rho _{ab\text{ }}^{\ast }$for YBCO$_{6.6\,}$.\ For YBCO$_{7}$\ the
deviation from linearity only occured for $\Delta \rho _{ab}>200\mu \Omega
.cm$.\ The corresponding correction, which did not exceed 10\% has not been
performed in fig.2.\FRAME{ftbpFU}{9.0325cm}{6.0231cm}{0pt}{\Qcb{Decrease of
the critical temperature $\Delta T_{c}$ as a function of the defect content $%
x_{d}$. The latter is proportional to $\Delta \protect\rho _{ab}$ measured
at 150K for two crystals of YBCO$_{7}$. In the case of YBCO$_{6.6}$ the data
(full circles) for $\Delta \protect\rho _{ab}$ has to be corrected into $%
\Delta \protect\rho _{ab}^{\ast }$ (full diamonds) which is a best estimate
of $x_{d}$ as explained in the text.\ The AG curve (dotted line) is
represented in the top part.\ The best linear fits (full lines) are also
diplayed together with those for EK shown in fig.4 using an analytical fit
for $\protect\rho (T_{c})$ versus $\Delta \protect\rho _{ab}$ or $\Delta 
\protect\rho _{ab}^{\ast }$(dashed lines).}}{\Qlb{Fig.2}}{Figure}{\special%
{language "Scientific Word";type "GRAPHIC";maintain-aspect-ratio
TRUE;display "USEDEF";valid_file "T";width 9.0325cm;height 6.0231cm;depth
0pt;original-width 597.25pt;original-height 398.5pt;cropleft "0";croptop
"1";cropright "1.0045";cropbottom "0";tempfilename
'H90XDV00.wmf';tempfile-properties "XPR";}}

\textit{The linear variation of }$T_{c}$\textit{\ with defect content, down
to} $T_{c}=0,$ is the most striking feature of the data displayed in fig.2.\
This result contrasts with that expected from the standard Abrikosov-Gorkov
formula which for d-wave superconductors gives $T_{c}$ as \cite%
{Borkowski,Sun}:

\begin{equation}
-\ln (\frac{T_{c}}{T_{c0}})\,=\Psi (\alpha +\frac{1}{2})-\Psi (\frac{1}{2})
\label{Eq1}
\end{equation}%
where $\Psi (x)$ is the digamma function , $\alpha =\hbar /(2\pi
k_{B}T_{c}\tau )$ is the pair-breaking parameter, and \ $1/\tau \varpropto
x_{d}$ is the scattering rate in the normal state . The scale of the $T_{c}$
reduction is fixed by the ratio of $\hbar /\tau $ and $k_{B}T_{c}$ so that
the impurity pair breaking becomes more and more efficient when $T_{c}$
decreases. This yields the well known negative curvature of the AG curve
shown in fig 2, \textit{which is obviously not observed in the present data }%
\cite{footnote5}. Some published data for impurity substitutions \cite%
{Tallon}\ have been fitted with Eq.\ref{Eq1}. Within the limited accuracy on
the impurity content and substitution site they could be fitted as well with
a linear variation.

Let us also consider the width of the superconducting transition, which
should reflect the inhomogeneities of the sample.\ The initial transition
width $\delta T_{c}\simeq 0.7K$ increases roughly linearly with $x_{d}$ but
never exceeds $5K$, and then decreases when $T_{c}$ approaches zero, as can
be seen in fig.3.\ \ Inhomogeneities due to the irradiation process, such as
electron energy losses through the sample, or a divergence of the electron
beam at the level of the sample, etc... result in a macroscopic distribution
of $x_{d}$ with the full width $\delta x_{d}=ax_{d}$. Within the AG theory
the transition width $\delta T_{c}$ should scale with the derivative of the
AG function.\ A large increase of $\delta T_{c}$ would then be expected when 
$T_{c}$ reaches zero, contrary to the experimental observation.\ From the
observed quasi linear variation of $\Delta T_{c}$ with defect content one
should expect a linear increase of $\delta T_{c}$ down to $T_{c}=0$. The
fact that $\delta T_{c}$ goes through a maximum rather indicates that
superconductivity becomes less sensitive to the distribution of defect
content.\ Such a possibility will be further considered in the discussion
hereafter.\ 

\FRAME{ftbphFU}{9.0325cm}{7.027cm}{0pt}{\Qcb{Width of the superconducting $%
\protect\delta T_{c}$ defined as shown in the insert as a function of $%
\Delta \protect\rho _{ab}$ for the two crystals of YBa$_{2}$Cu$_{3}$O$_{7}$.
The maximum of $\protect\delta T_{c}$ corresponds to a $T_{c}$ of $\sim 40K$%
. The variation expected from AG, EK and linear dependence of $T_{c}$ with $%
\protect\delta x_{d}/x_{d}\sim 7\%$ are also plotted (see text)}}{\Qlb{Fig.3}%
}{Figure}{\special{language "Scientific Word";type
"GRAPHIC";maintain-aspect-ratio TRUE;display "USEDEF";valid_file "T";width
9.0325cm;height 7.027cm;depth 0pt;original-width 483.8125pt;original-height
398.5pt;cropleft "0";croptop "1";cropright "1.0614";cropbottom
"0";tempfilename 'H90IP704.wmf';tempfile-properties "XPR";}}

Although AG theory is not applicable when localization corrections to the
conductivity occur, the experimental deviations are already evident for a
range of $T_{c}$ values for which these effects are negligible. A
quantitative discrepancy with AG\ theory has been also evidenced, as the
initial decrease of $T_{c}$\ is much slower than predicted \cite%
{Nachumi,Giapin,Tolpygo}.\ For YBCO$_{7}\;$for instance, theoretical
estimates of $\Delta T_{c}/\Delta \rho _{ab}$\ \cite{Radtke} range from $%
0.7\,$to$\,1.2\,K/\mu \Omega .cm$, a factor 2 to 4 higher than the
experimental value found here $(\sim 0.35\,K/\mu \Omega .cm)$. It has been
shown that this difference can be taken into account in the framework of the
AG model for a d-wave order parameter if one assumes an anisotropic impurity
scattering \cite{Haran}. With appropriate parameters, one can explain the
initial decrease of $T_{c}$ \textit{but with a similar dependence with }$%
x_{d}$\textit{\ as the AG curve, which does not solve the contradiction with
our results.\ }

A major reason for a breakdown of the AG theory might result from the very
short coherence lengths in the high $T_{c}$ cuprates, which does not allow
to assume a uniform gap averaged over the disorder. This might allow to
explain as well the variation of the superfluid density $n_{s}$ with $x_{d}$
as proposed by Franz et al \cite{Franz}.\ Such a possibility is reinforced
by the recent STM data \cite{STM-Zn} which reveal that the large depression
of the DOS\ peaks, and therefore of $n_{s}$, found nearby Zn impurities
disappears on a lengthscale comparable to the coherence length $\xi $. As
these regions of depressed superconductivity overlap for large $x_{d}$, a
simplistic all or nothing model ( so called ''swiss-cheese'') might explain
the negative curvature of $n_{s}(x_{d})$ which has been noticed by various
authors \cite{Nachumi,Bernhard}.\ Let us point out at this stage that the
observation in underdoped pure compounds of a linear relation between $T_{c}$
and $n_{s}$ \cite{Uemura}, has led some authors to consider that this
relation applies as well for impure samples \cite{Ghosal,Morais}.\ The
present observation of a linear decrease of $T_{c}$ \textit{definitely
contradicts this simple guess}.

In a totally different approach, Emery and Kivelson (EK) \cite{Emery1} argue
that in low $n_{s}$ superconductors, $T_{c}$ might be determined by phase
fluctuations of the order parameter. The temperature of the classical phase
ordering $T_{\theta }^{\max }$ which is proportional to $n_{s}/m^{\ast }$
can be much lower than the mean field critical temperature and is therefore
an upper bound on the true $T_{c}$. In that case the influence of disorder
is to increase quantum phase fluctuations.\ They propose \cite{Emery2} that
their magnitude is determined by the value of $\rho (T_{c}),\;$and that
superconductivty disappears for a critical value $\rho _{Q}$, so that

\begin{equation}
\ln (T_{\theta }^{\max }/T_{c})=\rho (T_{c})/\rho _{Q}\ln (\epsilon /T_{c})
\label{Eq.2}
\end{equation}%
in which $\epsilon $ is the energy scale of pairing interactions.

\FRAME{ftbphFU}{8.5339cm}{6.7239cm}{0pt}{\Qcb{The decrease of $T_{c}$ as a
function of $\protect\rho (T_{c})$ as defined in the insert of fig.3 is
compared to Eq.2.\ The best fits are obtained for $\protect\rho _{Q}$ equal
to $600$\ and $1380$ $\protect\mu \Omega .cm$ respectively for YBa$_{2}$Cu$%
_{3}$O$_{7}$ and YBa$_{2}$Cu$_{3}$O$_{6.6}$.}}{\Qlb{Fig.4}}{Figure}{\special%
{language "Scientific Word";type "GRAPHIC";maintain-aspect-ratio
TRUE;display "USEDEF";valid_file "T";width 8.5339cm;height 6.7239cm;depth
0pt;original-width 547.0625pt;original-height 410.5625pt;cropleft
"0";croptop "1.0476";cropright "1";cropbottom "0";tempfilename
'H8MBKD02.wmf';tempfile-properties "XPR";}}

The fact that we have access in our experiment to $T_{c}$ values near zero
allows us to test this dependence very precisely in fig.4, where the data
for $T_{c}$ are plotted versus \ $\rho (T_{c}).\;$Reasonable fits of our
data can be obtained with a large range of $\epsilon $ values. As suggested
by EK, we have therefore chosen a realistic value $\epsilon \thicksim 1200K$%
, the antiferromagnetic exchange energy in YBCO$_{7}$. As can be seen in
fig.4, the data can be well fitted by Eq.2 with $T_{\theta }^{\max }=103K$
and $\rho _{Q}=600\mu \Omega .cm$ for YBCO$_{7}$ ($76.6K$ and $1380\mu
\Omega .cm\;$for YBCO$_{6.6}$). Whatever the value chosen for $\epsilon $ we
always found a value of $\rho _{Q}$ roughly two times larger in YBCO$_{6.6}$
than in YBCO$_{7}$ as it is mainly given by the value of $\rho (T_{c})$
corresponding to $T_{c}=0$.

We can as well determine the expected evolution of the transition width $%
\delta T_{c\text{ }}$assuming $\delta x_{d}=ax_{d}.\;$We have done this for
YBCO$_{7},$ taking the derivative of\ Eq.\ref{Eq.2} with an analytical fit
of the data for $\rho (T_{c})$ versus $\Delta \rho _{ab}$.\ The variation of 
$\delta T_{c}$\ calculated with the parameters deduced from the fit of Fig.4
reproduces rather well the trend of the experimental data in Fig. 3 with $%
a=0.07$. This analysis can conclusively be done solely from the data near $%
T_{c}=0$ and\ allows to explain jointly the variation of $T_{c}$ and of $%
\delta T_{c}$, while all the alternative possibilities examined so far did
not. So this emphasizes the importance of quantum phase fluctuations$\ $and
of\ $\rho (T_{c})$ in determining the actual $T_{c}$ in this highly damaged
region. Let us point out that the values for $\rho _{Q}$ taken per CuO$_{2}$
sheet \cite{footnote6} ($\rho _{Q}^{2D}=10k\Omega /\square $ and $25k\Omega
/\square $ for YBCO$_{7}$ and YBCO$_{6.6}$) are much smaller than those
observed in ion irradiated thin films (ref.12 in \cite{Emery2}).\ This is
probably related to the fact that irradiation damage in thin films usually
induces much larger increase of resistivity than in single crystals. However
the large variation of $\rho _{Q}^{2D}$ with hole doping found here is not
anticipated by EK. This might stem from the fact that the analysis of EK is
a priori more appropriate to describe underdoped cuprates with low
superfluid density. One expects a behaviour completely different for
overdoped materials for which T$_{c}$ should be determined by the mean-field
transition temperature and disorder should be mainly pair breaking.

This leads us as well to consider that our fits are not a proof that the
analysis of EK applies for the initial decrease of $T_{c}$ particularly in
optimally doped YBCO$_{7}$ which is at the borderline\ between underdoped
and overdoped behaviour. It is interesting at this stage to remind the
results obtained for the initial decrease of $T_{c}$ in cuprates with
different hole dopings $n_{h}$. We have shown that $\Delta T_{c}/\Delta \rho
_{ab}^{2D}$ increases steadily with $n_{h}$ \cite{RA}.\ Obviously the
initial slope of Eq.\ref{Eq.2}, that is $\Delta T_{c}/\Delta \rho
=-(T_{c}/\rho _{Q})$ln$(\epsilon /T_{c})$ cannot explain this behaviour
unless one introduces a large non physical reduction of $\rho _{Q}$\ with
increasing hole content. If the number of carriers is taken as the number of
holes $n_{h}$, the initial slope is such that $\Delta T_{c}$ scales as $%
m^{\ast }/\tau $ all over the phase diagram\ \cite{RA}, which indicates that
pair breaking in the d-wave superconducting state might still contribute to
the initial decrease of $T_{c}$.\ This possibility is favored by the STM
observation, in optimally doped Bi-2212, of quasiparticle states around Zn
impurities, as expected from strong scattering theories \cite{STM-Zn}. A
reasonable explanation of the data would then be the occurence of a
crossover from pair breaking to phase fluctuations regime for high disorder.
Such a crossover should progressively shift towards increasing disorder with
increasing doping.

In conclusion the possibility to create very homogeneous disorder and well
controlled defect contents $x_{d}$ has allowed us to study carefully the
depression of $T_{c}$ down to $T_{c}=0$ in underdoped and optimally doped
YBCO. We have shown that $T_{c}$ which decreases quasi linearly with $x_{d}$%
, does not scale with the reported variation of $n_{s}.$The observation of
the narrow transition width in YBCO$_{7\text{ }}$for large $x_{d}$ allowed
us to suggest that $\rho (T_{c})$ is the relevant scattering rate which
controls $T_{c}$ for high damage, as expected from quantum phase
fluctuations. The present experiments provide sufficiently detailed
information to stimulate further theoretical calculations which should take
into account both the amplitude and phase variation of the order parameter.

As for the highly damaged samples,we might anticipate that their
superconducting properties should strongly differ from those observed in
weakly disordered compounds. Although this regime occurs near superconductor
to insulator transition, further work is clearly needed to understand the
relationship with the superconductor-insulator and the metal-insulator
transition observed in other 2D structures.


\begin{thebibliography}{99}
\bibitem{footnote1} The error in the absolute values of $\rho _{ab}$ ( 5\%)
is determined by that on the sample thickness. To compare the data on the
two samples, we have therefore chosen to normalize the data to equalize the
slopes of the $\rho (T)$\ curves at high $T$ for the pure samples.\ This
corresponds to a correction of 3\% on sample\#2.
\bibitem{footnote2} We can estimate the energy loss of 2.5MeV electrons to
be $\sim $ $1keV/\mu m$ in YBCO.
\bibitem{footnote3} We have previously shown that 1\% of planar defects
induce a reduction in $T_{c}$ of $10K$ in YBCO$_{7}$ \cite{RA}.
\bibitem{footnote4} As long as the defect content \ is low ($<10\%$), one
can assume that the defect concentration $x_{d}$ is proportional to the
electron fluence.
\bibitem{footnote5} If we take into account the fact that $1/\tau $
increases for large $x_{d}$ at low T, this would lead to an even larger
deviation of Eq.1 with experimental data.
\bibitem{footnote6} The 2D sheet resistance $\rho ^{_{2D}}$ is determined as 
$\rho ^{_{2D}}=n.\rho _{ab}/c$ where $c$ is the $c$-axis parameter and $n$
is the number of CuO$_{2}$ planes per unit cell. The values found here have
an order of magnitude comparable to the value $h/4e^{2}=6.45k\Omega /\square 
$ usually quoted for the superconducting to insulator transition in 2D
metals (\cite{SI transition}). One can notice in fig.2 that $T_{c\text{ }}$%
vanishes for\ both hole contents for $\Delta \rho _{ab}^{2D}$ of the order
of $6.4k\Omega /\square $ as also seen for Zn substitution \cite{Fuku},
although $\rho (T_{c})$ is much larger.
\end{thebibliography}
\end{document}